\documentclass[10pt]{iopart}
\usepackage{graphicx}
\usepackage{iopams} 
\eqnobysec
\begin{document}
\date{\today}
\title{How to control nonlinear effects in Binder cumulants}

\author{Y. Meurice
}

\address{Department of Physics and Astronomy\\ The University of Iowa\\
Iowa City, IA 52242 USA\\
\vskip10pt
email: yannick-meurice@uiowa.edu}




\begin{abstract}
We point out that ignoring nonlinear effects in finite size scaling may lead to errors in estimates 
of the critical temperature and Binder cumulants. We show that the order of magnitude of these effects can be estimated from data at relatively small volume. Using this estimate, we propose to use linear 
fits in increasingly small temperature regions as the volume is increased (rather than using a fixed temperature interval). The choice of the exact coefficient of proportionality can be optimized and reveals interesting crossing patterns among estimates. 
We show that the new procedure works very well for Dyson' hierarchical model. We discuss applications of the method for 
3 dimensional spin models and finite temperature lattice gauge theories and comment on the nonlinear effects for existing calculations. 

\end{abstract}

\section{Introduction}

The study of the Binder cumulants for isolated systems of various sizes has been an important tool to determine the critical temperature and universal quantities of spin models 
\cite{binder81,barber85,ferrenberg91,olsson96,blote96,gupta96,hasenbusch98,campostrini99,binder01} and lattice gauge theory models at finite temperature \cite{engels89,fingberg92,sinclair06,sinclair07,deforcrand07a,deforcrand07b,velytsky07}. 
The basic idea is that the intersections among Binder cumulants curves for systems of different sizes obey scaling relations often referred to as Finite Size 
Scaling (FSS). 

Linearization in the scaling variables is often used to derive approximate FSS relations 
for the Binder cumulants and the critical temperature that allow, in principle, simple linear extrapolations to infinite volume \cite{binder81,ferrenberg91}. 
However, in practice, the errors bars on the intersections are often of the same order as the finite 
size variations and the linearity of the data is not always obvious. Examples of such graphs are figure 7 in  \cite{ferrenberg91} or figures 3, 4 and 6 of  \cite{velytsky07}. For this reason, authors will sometimes prefer to rely on a single intersection for the two largest volumes available that has potentially smaller errors \cite{gupta96,velytsky07}. 

A possible source of systematic error in these infinite volume extrapolations are the nonlinear terms that have been neglected in the derivation of the linear FSS formulas. If these effects are small enough, it is difficult to disentangle them from the statistical errors. Typically, we are talking 
about a 1-2 percent effect. This may sound small, however, for the $O(N)$ hierarchical sigma models
where very accurate calculations can be performed, the differences between the Binder cumulants at successive values of $N$ are less than 2 percent for 
low values of $N$ \cite{moprogress}. Consequently, it is reasonable to set accuracy goals of less than 1 percent for the 
Binder cumulants, especially if the estimate is used to establish that a particular model 
belongs to a known class of universality. 

In this article, we discuss quantitatively the effects of the nonlinear terms in FSS with an example and we propose a new 
procedure to obtain infinite volume estimates that are minimally distorted by the nonlinear effects. 
The procedure can be used for spin models or lattice gauge models and $\beta$ is a notation that can be used interchangeably for either the inverse temperature in Ising-like models or $2N_c/g^2$ in gauge theory.

One purpose of the article is to provide a specific improvement to the common practice of using data at the same, equally spaced, values of $\beta$ for all value of the linear size $N$ (we call this the ``fixed interval procedure"). Clear examples of the fixed interval procedure can be seen 
in figure 6 of \cite{barber85}, figure 11 of \cite{olsson96}, figure 7 of \cite{engels89}, figure 1 of 
\cite{fingberg92} and figure 1 
of \cite{velytsky07} where the fourth order cumulant is plotted versus $\beta$ or the temperature. 
In the large volume limit, the cumulant becomes a function (that we will call $f$) of the reduced variable $((\beta-\beta_c)/\beta_c)N^{1/\nu}$, with $\nu$ the usual correlation length exponent. 
This function $f$ can be approached at sufficiently large volume by plotting the cumulant versus this reduced variable. When this is done, the data ``collapses'' up to some finite size corrections. 
It seems clear that in order to reduce the nonlinear effects, we should shrink the range of $\beta$ 
as $N$ increases. 

This idea seems simple, however its practical implementation is non-trivial. First, we need to estimate the size of the nonlinear effects. Second, we only know 
$\beta_c$ with a finite accuracy and the interval cannot be shrunk to a size 
much smaller than the uncertainties on $\beta_c$. 
We need to find a compromise between two conflicting requirements. On one hand, the interval needs to 
be small enough to reduce the nonlinear effect. On the other hand, if the interval is too small, we may 
not cover properly the region where the linear approximation is valid which could result in errors for the intersections. 
The main technical question that we address here is: can we find an optimal width for the interval 
were both requirements can be reasonably satisfied? 
It is difficult to answer this question for the gauge or spin models discussed above because
it is hard to extract the nonlinear effects from the statistical errors. In addition, 
we have no independent estimates of the infinite volume cumulants 
and calculations at large volume are CPU demanding. 

We will address this question in a model where the practical 
issues mentioned above can be avoided, namely 
Dyson hierarchical model \cite{dyson69} with an Ising measure. In this model, it possible to block spin 
very accurately with 
methods \cite{guide,gam3rapid,gam3} that have been recently reviewed in  \cite{hmreview}. This allows us to make very accurate calculations at very large volume 
(linear size up to $10^6$ sites) that cannot be reached 
with ordinary Monte Carlo methods. 
The free parameter of this model has been fixed in such a way that free Gaussian fields scale like in 3 dimensions. The technical details are given in \ref{sec:appa}. These large volume numerical calculations will be used 
later to check the accuracy of estimates made using data at much smaller volumes (linear size up to 256 sites as in existing Monte Carlo (MC) calculations).

Our presentation is focused on the estimation of the infinite volume critical temperature and the (universal) fourth order Binder cumulant. This quantity is defined precisely in 
section \ref{sec:notations} where we introduce other notations and state the  problem. We provide a 
parametrization of the nonlinear effects for the fourth order cumulant in terms of two new parameters 
(see equation (\ref{eq:param})).
Accurate numerical values of these parameters 
are given in \ref{sec:appa} in the case of the 
hierarchical model. We then discuss the linearization and graphs testing its validity. 

The new method that we propose here (the ``shrinking interval procedure") 
is presented in section \ref{sec:shrink}. It 
proceeds in two steps. First, 
we need to estimate the 
size of the nonlinear effects using data at relatively small volume. We explain that the important 
quantity is the ratio of the linear to quadratic leading amplitudes (called $f_1/f_2$).
We show in \ref{sec:pencil} that this quantity can be estimated with reasonable accuracy
using a pencil and a ruler on a graph showing the  
Binder cumulant versus $\beta$ at relatively small volume. 
We then propose to use a shrinking range $\Delta \beta = \epsilon\beta_c (f_1/f_2)N^{-1/\nu}$ where $\epsilon$ 
is an adjustable small parameter which we shall see, can picked in an optimal way by localizing crossings in infinite volume extrapolations.

In section \ref{sec:compare} we compare the shrinking interval procedure with the fixed interval procedure. The calculations are done using a number of values of $\beta$ and linear sizes that 
are typical in existing MC calculations. 
We emphasize that in doing these calculations, we did not use our prior knowledge of the critical 
temperature, Binder cumulant or amplitudes given \ref{sec:appa}, however we used our knowledge of the 
critical exponents $\nu$ and $\omega$. We first used a fixed value of $\epsilon$ (0.05) and found much better result than with the fixed interval method. We then compared infinite volume extrapolations 
for $\beta_c$ and the cumulant, 
for different values of $\epsilon$. The results are displayed in figure \ref{fig:extraps} where we see 
remarkable crossings (among extrapolation curves for different maximal volume) at approximately the same value of $\epsilon$ for the two quantities. 
We compared the results for this optimal value with the accurate ones and found relative errors of less than $10^{-5}$ for $\beta_c$ and of less than $5\times 10^{-3}$ for the cumulant. 

Figure \ref{fig:extraps} is the most important result of the paper. It shows that an optimal value 
of $\epsilon$ can selected using the numerical results. It is crucial to realize that the crossing 
means that for the optimal value of $\epsilon$, results at not too large volume provide very accurate results. One should also appreciate the number of large volume calculations involved in producing the 
two graphs that would take a prohibitively long time for 4 dimensional lattice gauge theory. 

In section \ref{sec:app}, we discuss the effects of the nonlinear terms for the calculations done 
in the literature and we conclude with possible applications of the new method. 
\section{Statement of the problem and notations}
\label{sec:notations}
The new method proposed in this article can be used for two types of models. 
First, the spin models where $\beta$ denotes the 
inverse temperature in appropriate units. In the literature on the 3 dimensional Ising model \cite{binder01} 
$\beta$ is often denoted $K$. 
Second, the lattice gauge models with gauge group $SU(N_c)$ at finite temperature $T=(N_\tau a)^{-1}$ where $N_\tau$ is the number of sites in the 
Euclidean time direction and $a$ the lattice spacing. 
For these models $\beta=2N_c/g^2$ is not the inverse temperature. However, if we use the one-loop scaling \cite{fingberg92}, 
\begin{equation}
(T-T_c)/T_c \simeq (\beta -\beta_c)12\pi^2/11N_c^2 \ .
\end{equation}
This formula illustrates that for the finite temperature gauge model, $T>T_c$ implies $\beta >\beta_c$ because the ordered phase of these models corresponds to high $T$ and 
high $\beta$. 

In summary, the notation $\beta$ can be used for both types of models with a consistent meaning: the ordered phase corresponds to $\beta>\beta_c$. For this reason, we will define a reduced $\beta$ (or $K$) 
variable: 
\begin{equation}
	\kappa \equiv (\beta-\beta_c)/\beta_c
\end{equation}
Other choices such as $(\beta-\beta_c)/\beta $ lead to similar linear behavior 
near $\beta_c$ but have different nonlinear behavior. These considerations are 
important if we want to connect with other expansions \cite{campbell06,campbell07}.

We now consider finite size scaling for isolated blocks in 3 dimensions. The spin models are defined on 
symmetric cubic lattice $N^3$ sites. The gauge models are defined $N_\tau\times N_\sigma^3$ lattices. 
We use the convention $N_\sigma=N$ for the variable number of sites (while $N_\tau$ is kept fixed) in order to have unified notations.  In both cases, the system is not in contact with a larger system but isolated 
and defined with suitable boundary conditions. 

Under a RG transformation where the lattice spacing $a\rightarrow \ell a$, we 
have $N\rightarrow N/\ell$ and $u_i\rightarrow\ell^{y_i} u_i\ $, 
where the $u_i$ are the nonlinear scaling variables which transform multiplicatively. 
We denote $u_{\kappa}$ the only relevant scaling variable (we will not deal with external fields) and 
we will only consider the effect of the first irrelevant variable denoted $u_1$.
We assume that the exponents $y_{\kappa}=1/\nu $,  
	$y_1=-\Delta/\nu=-\omega $ 
are known  with good precision as it is the case when we are testing the hypothesis that a particular 
system belongs to a well studied class of universality.
In addition, we assume the expansions
\begin{eqnarray}
	u_{\kappa}&=&\kappa+u_{\kappa}^{(2)}\kappa^2+\dots\\
	u_1&=&u_1^{(0)}+u_1^{(1)}\kappa +\dots
\end{eqnarray}

\begin{figure}
\includegraphics[width=2.4in,angle=270]{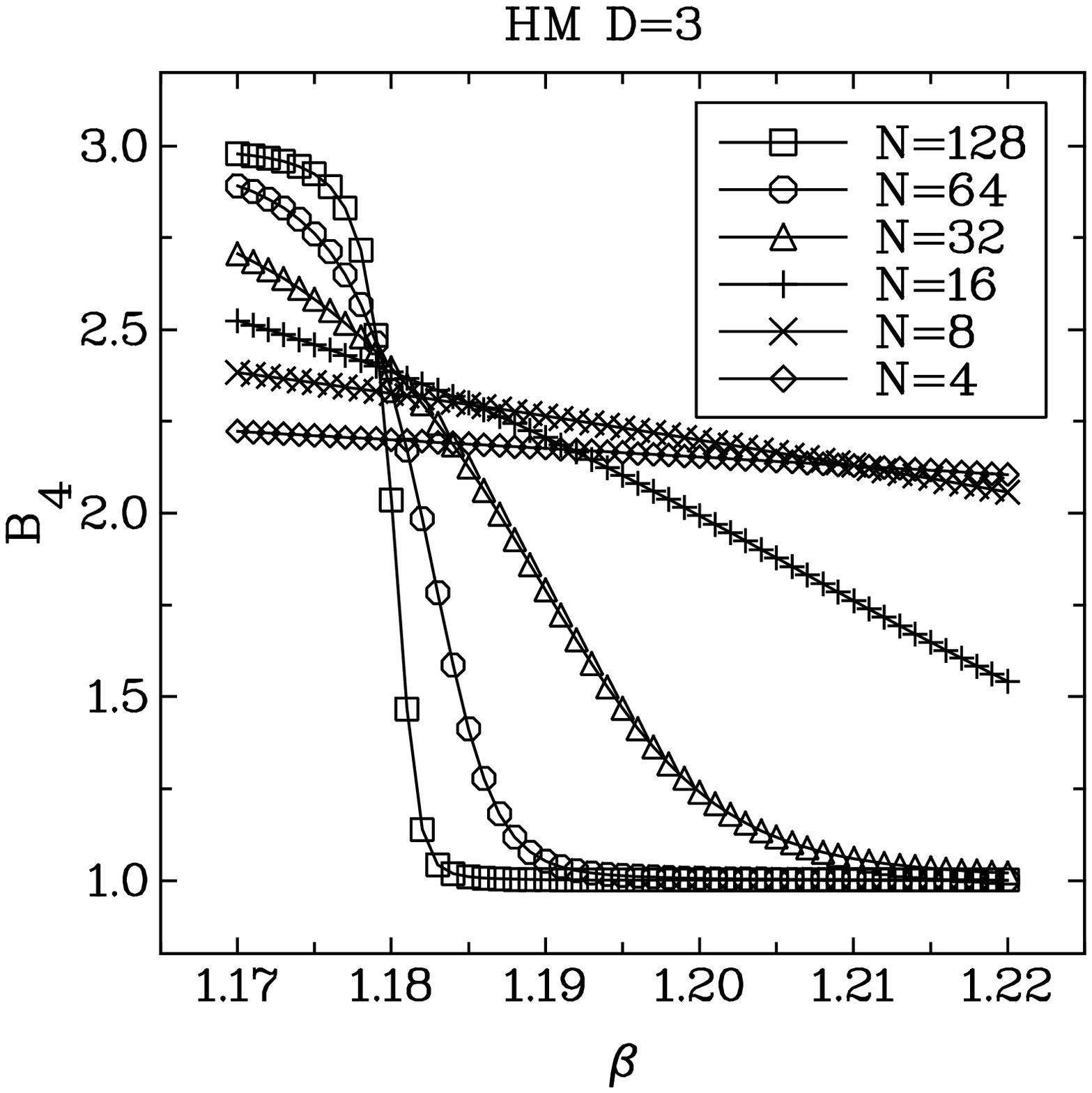}
\includegraphics[width=2.4in,angle=270]{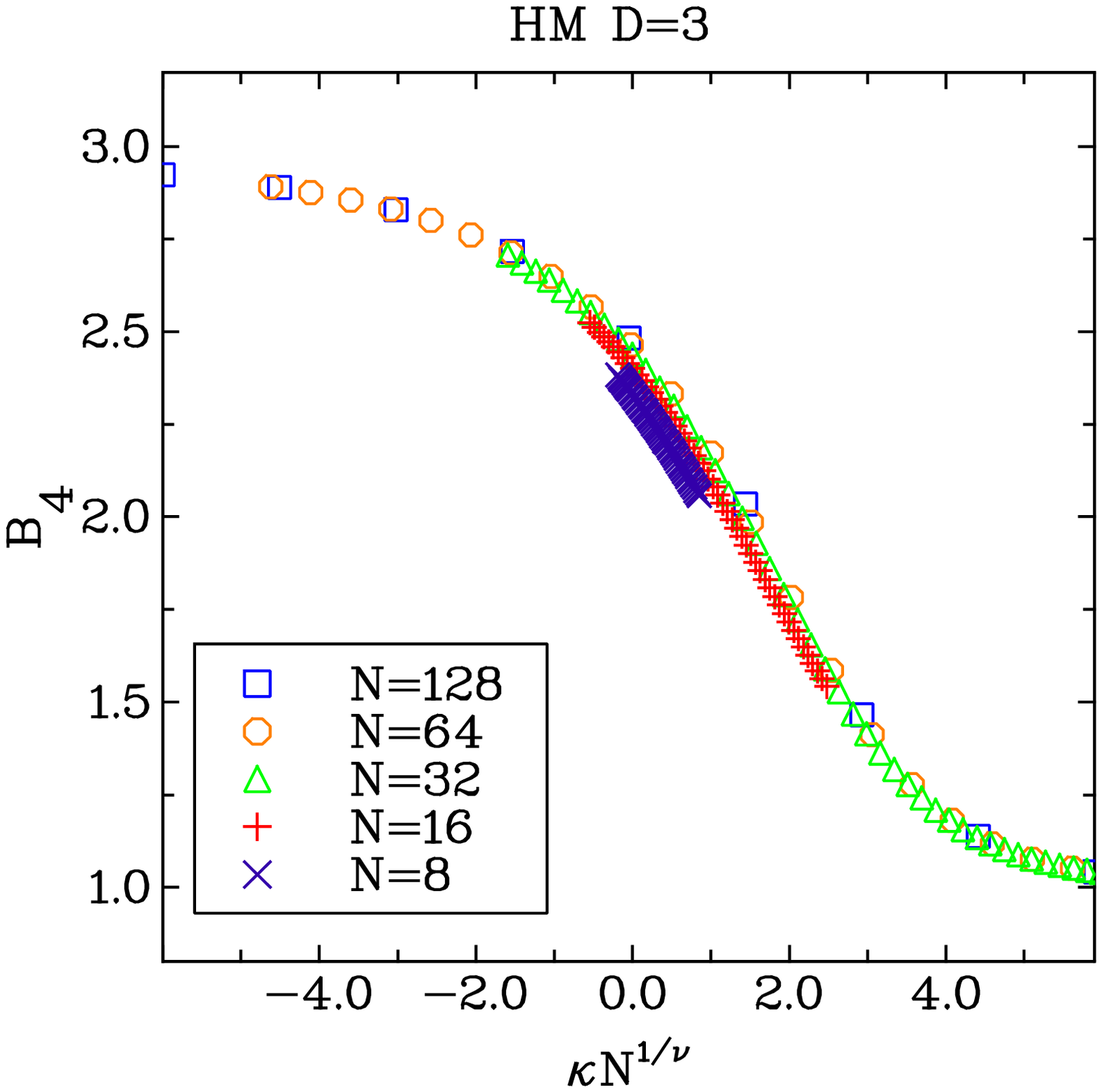}
\caption{$B_4$ versus $\beta$ (left) and versus $\kappa N^{1/\nu}$ (right) , for $N=$ 4, 8, 16, 32, 64 and 128,  for the Ising hierarchical model.}
\label{fig:b4}
\end{figure}

We define the fourth Binder cumulant as 
\begin{equation}
	B_4\equiv <m^4>/<m^2>^2
\end{equation}
with $m$ the average of the order variable (spin for Ising, Polyakov's loop in the time direction for gauge theory) in 3 dimensions. 
Directly related quantities appear in the literature such as $U=1-B_4/3$ or $Q=1/B_4$ etc.
Due to the lack of consensus, we have picked a form that has limits that are easy to remember. 
Both numerator and denominator are unsubtracted averages and so this 
quantity can be calculated at large volume without running into loss of accuracy problems 
occurring for subtracted averages. In the symmetric phase, $<m^4>$ is dominated by $3<m^2>^2$ and the 
the correction is suppressed by one power of the volume (the coefficient is a renormalized coupling constant).
Consequently, if $\beta<\beta_c$ ($\kappa < 0$), $B_4$ tends to 3 in the infinite volume limit. On the other hand in the ordered phase, $<m^4>$ is dominated by $<m^2>^2$ and the limit is 1. 
At fixed $N$, for sufficiently small $\beta$, $B_4$ is close to 3 and for sufficiently large $\beta$, 
$B_4$ is close to 1. As $N$ increases, the transition sharpens as illustrated in figure \ref{fig:b4}.
Finite size scaling \cite{binder81,binder01}
implies that 
\begin{equation}
B_4=f(u_{\kappa} N^{y_{\kappa}},u_1 N^{y_1},\dots)
\label{eq:exactfss}
\end{equation}
Since $y_1<0$, when the volume is large enough, $B_4$ can be approximated by a function of  
$\kappa N^{y_{\kappa}}$ only. If we plot $B_4$ versus $\kappa N^{y_{\kappa}}$, the curves at various $N$ approximately collapse. This is illustrated in figure \ref{fig:b4} where one can see that the larger violations of this approximation are observed at smaller volume. 
The most common procedure to study the intersections of the various curves at fixed $N$ is to linearize equation (\ref{eq:exactfss}) and try to extrapolate the results at infinite $N$. 
In the rest of this article, we will consider the effects of nonlinear terms on this procedure. 
Our main assumption will be that 
\begin{equation}
\hskip-30pt
B_4(\beta, N)\simeq B_4(\beta_c,\infty)+f_1 \kappa N^{1/\nu}+f_2\kappa^2 N^{2/\nu}+ (c_0+c_1\kappa N^{1/\nu})N^{-\omega}\ .
\label{eq:param}
\end{equation}
Note that we have not included terms of the form $\kappa^2 N^{1/\nu}$ which should appear as a consequence of the presence of $\kappa^2$ terms in the expansion of $u_{\kappa}$ because they 
are suppressed by a factor $N^{-1/\nu}$ compared to the term $(\kappa N^{1/\nu})^2$. 
We have also not included terms of order $N^{-2\omega}$ and $N^{-|y_2|}$.

In the linear approximation ($f_2=c_1=0$), we recover the standard linear FSS formula for the point of  intersection denoted $(\beta^{\star}(N,N'),B_4^{\star}(N,N'))$
between the two curves $B_4(\beta,N)$ and $B_4(\beta,N')$, namely
\begin{eqnarray}
	\beta^{\star}(N,N')&=&\beta_c+\beta_c(c_0/f_1)L(N,N')\ , \nonumber \\
	B_4^{\star}(N,N')&=&B_4+c_0M(N,N')\ ,
	\label{eq:linearfss}
\end{eqnarray}
with 
\begin{eqnarray}
L(N,N')&=&	(N^{-\omega}-N'^{-\omega})/(N'^{1/\nu}-N^{1/\nu})\ ,\nonumber \\
M(N,N')&=&(N^{-\omega-1/\nu}-N'^{-\omega-1/\nu})/(N'^{1/\nu}-N^{1/\nu})\ .           
\label{eq:LM}
\end{eqnarray}
These formulas have been written in a way that makes the symmetry under the interchange of 
$N$ and $N'$ obvious. They may look more familiar if we replace $N'$ by $bN$ and factor out the 
powers of $N$. Since we assume that $\nu$ and $\omega$ are known in good 
approximation, linear fits can used to determine the remaining unknown 
parameters in equation (\ref{eq:linearfss}). Graphs based on these formulas can be found 
in \cite{ferrenberg91} (for $b$ = 2), \cite{velytsky07} and in figures \ref{fig:fixedbet} and \ref{fig:fixedbin} below.

For reasons explained in the introduction, we will make model calculations for Dyson's hierarchical 
model. For this model, the local potential approximation 
is exact and the change in the local measure under block spinning can be calculated very accurately 
at very large volume and very close to $\beta_c$ \cite{guide,gam3rapid,gam3} (see  \cite{hmreview} for a recent review). 
The free parameter of this model, usually denoted $c$ has been fixed in such a way that free Gaussian fields scale like in 3 dimensions.
With this choice, as the blockspinning reduces the number of sites by a factor 2, the linear sizes are of the form $N=2^{n/3}$. The details of the calculations and accurate estimates of the parameters 
entering in equation (\ref{eq:param}) are given in \ref{sec:appa}.

\section{The shrinking interval procedure}
\label{sec:shrink}
In this section, we present the shrinking interval procedure 
advocated in the introduction. It may be useful to state all the steps that need to be followed. First, we perform linear fits 
of $B_4$ as a function of $\beta$ at fixed volume in a sufficiently small $\beta$ interval.
We then use the linear fits at different volumes to determine the intersections called $\beta^{\star}(N,N')$ and $B_4^{\star}(N,N')$ in equation (\ref{eq:linearfss}). Finally, we select 
a set of pairs $(N,N')$ and perform linear fits to determine the unknown coefficients in the linear FSS formula equation (\ref{eq:linearfss}). 
The final result is an infinite volume extrapolation for $\beta_c$ and $B_4$. 

We now discuss the first step. We need to specify the $\beta$ interval (in other words, its center and width) for calculations of $B_4$ at a given volume.  
Given a linear size $N$, we propose to use our best estimate $\bar{\beta_c}$ of 
$\beta_c$ obtained from smaller sizes. This is an iterative procedure. We need to start with some reasonable estimate (for instance obtained with the finite interval procedure at small size) and then keep improving. We propose to restrict the calculation of $B_4$ to the 
interval parametrized in the following way:
\begin{equation}
|\beta - \bar{\beta_c}|<\epsilon (f_1/f_2)\bar{\beta_c} N^{-1/\nu} \ .
\end{equation}
We do not need a very accurate value for $f_1/f_2$. We show in \ref{sec:pencil} that it is easy to estimate the order of magnitude of  $f_1/f_2$ from a graph such as figure \ref{fig:b4}, using a pencil and a ruler. 
The value of $\epsilon$ needs to be chosen carefully. On one hand, we need $\epsilon$ small enough in order to 
control the nonlinear effects. On the other hand, if $\epsilon$ is very small, we need a correspondingly 
good estimate of $\beta_c$. In addition, when $\epsilon$ is too small, the intersections may be far away from the regions where we have values of $B_4$. These two effects could in principle compensate. 
Unfortunately, in the example where we have done accurate calculations, they go in opposite directions.
Namely, the values of $\bar{\beta_c}$ obtained from small volume data are 
below the true $\beta_c$ while the small volume intersections are above $\beta_c$ as can easily be 
seen from figure \ref{fig:b4}. 

We want to figure out if the two conflicting requirements can be partially satisfied for some 
optimally chosen value of $\epsilon$. The errors on $\beta^{\star}(N,N')$ and $B_4^{\star}(N,N')$ 
depend not only on $\epsilon$ but also on $\bar{\beta_c}$. The choice of $\epsilon$ is clearly 
a difficult optimization problem and it is useful to make numerical experiments.
In the following section we show that values of $\epsilon \sim 0.1$ lead to results that compare very well with the results of \ref{sec:appa} and that a reasonable compromise between the two requirements discussed above is possible. 

We should also mention that the use of linear fits to determine the intersections could be replaced 
by more sophisticated search methods. However, in calculations that are CPU demanding, and where 
we have only data for a few values of $\beta$, linear fits is the most common practice.

\section{Comparing the fixed and shrinking interval procedures}
\label{sec:compare}

In this section, we compare FSS results using the shrinking interval procedure discussed above and the 
fixed interval procedure where we use 
fixed values of $\beta$ for all values of $N$. 
The calculations were made for the Ising hierarchical model. 
For the fixed interval calculation, we took the same 9 values of $\beta$: 1.176, 1.177, ..., 1.184 for 
all possible $N$ between 8 and 256, or in other words, $9\leq n\leq 24$ with $n$ the number of blockspinnings. For the shrinking interval procedure, we started with the fixed interval data with $N\leq16$ and an estimate of $\beta_c \simeq 1.1772$ corresponding to this data. 
We then followed the procedure described in section \ref{sec:shrink} with $\epsilon=0.05$. 
This value of $\epsilon$ was selected from the compromise of having linear fits that were not blatantly 
off the data and at the same time keep most of the intersections within the fitted range. We also 
picked 9 values of $\beta$ within the specified range. 
At each $N$, we used all the possible intersections involving the 5 closest sizes to determine $\bar{\beta_c}$ using equation (\ref{eq:linearfss}). This means that we made 
linear fits with 15 points and the results are quite robust under small changes in the procedure. 

\begin{figure}
\includegraphics[width=2.4in,angle=270]{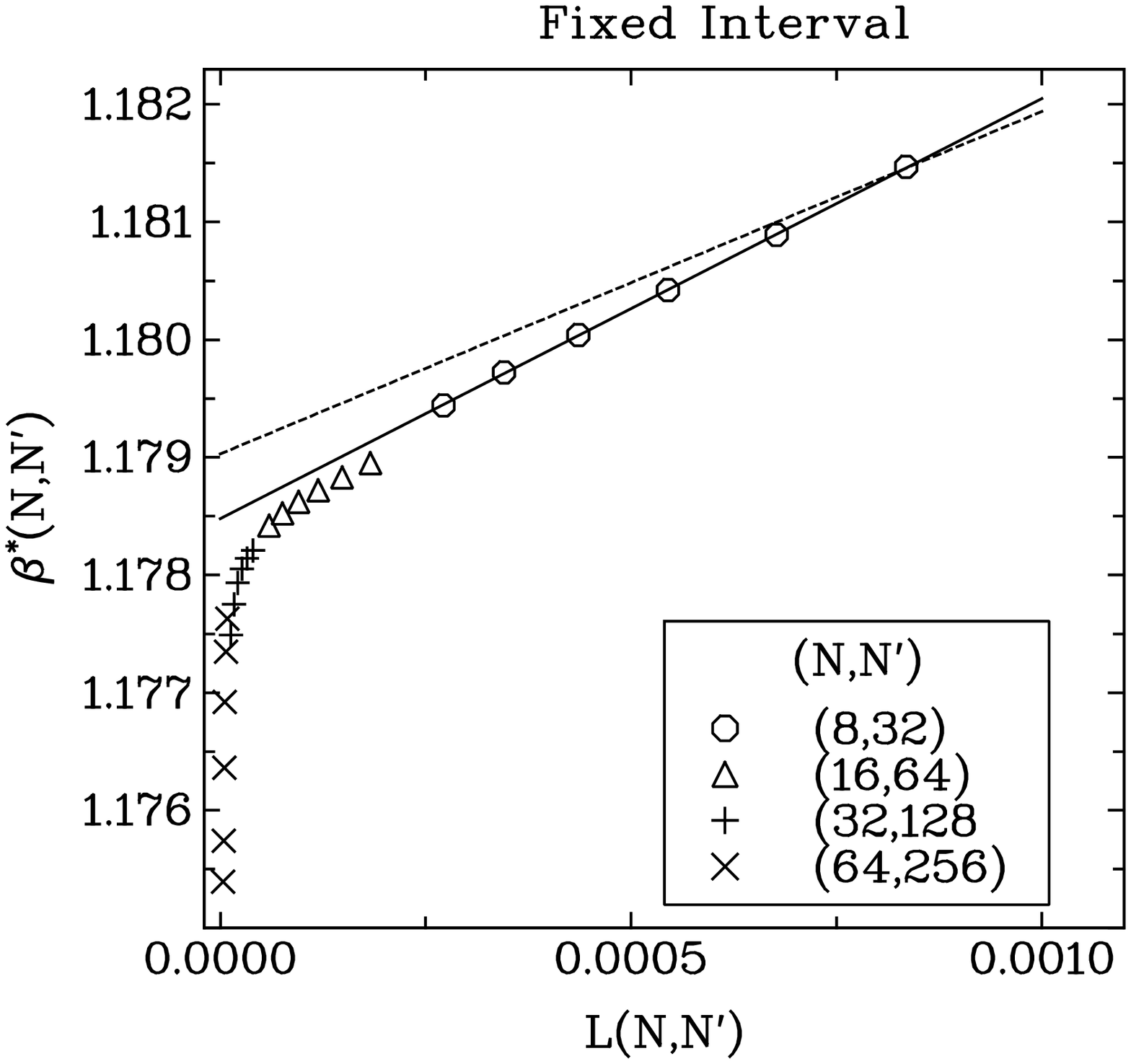}
\includegraphics[width=2.4in,angle=270]{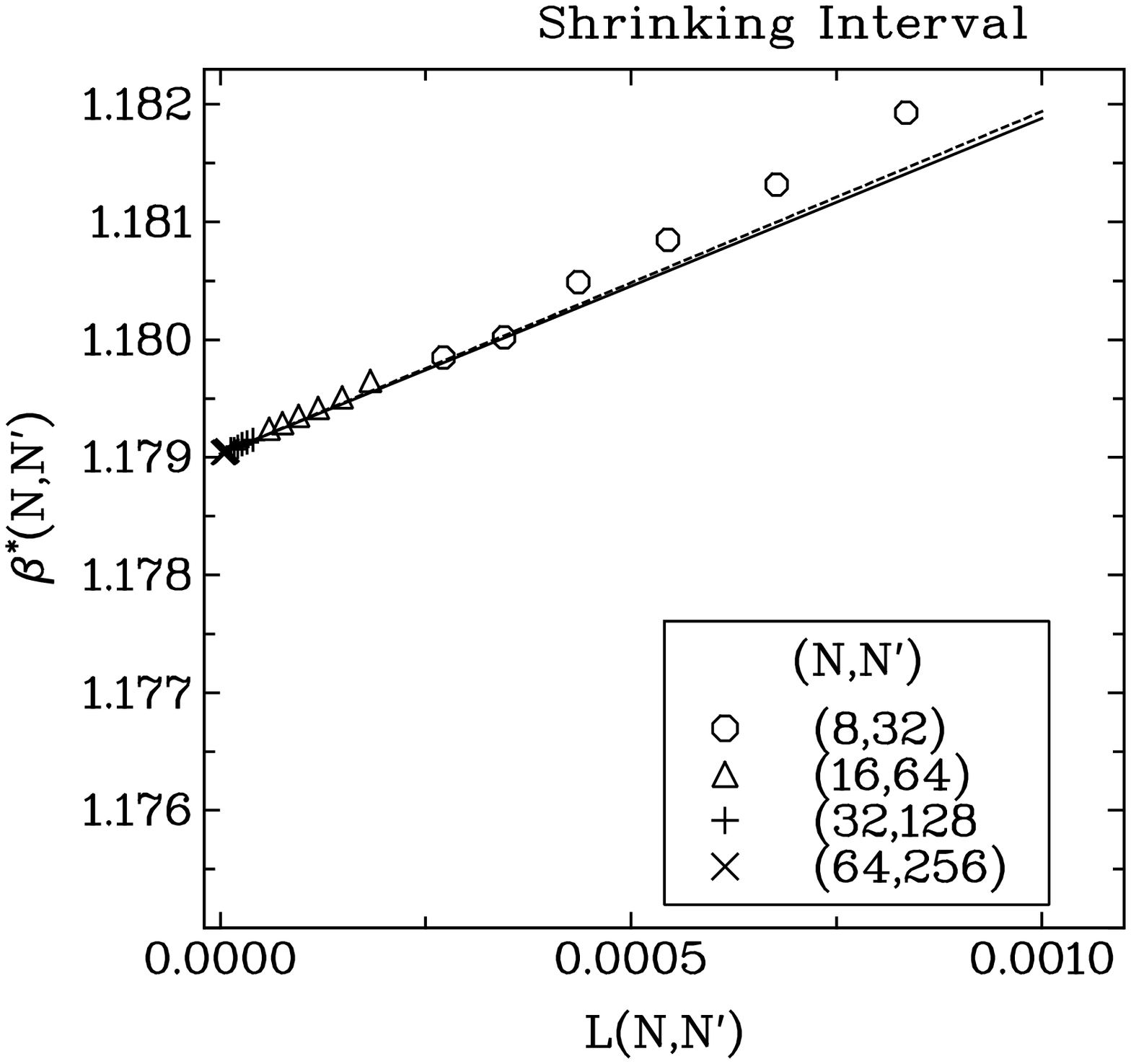}
\caption{Empirical values of $\beta^{\star}(N,N')$ obtained with the fixed interval procedure (left) and with the shrinking interval procedure (right) versus $L(N,N')$ for 4 sets of 6 pairs of values 
defined in the text. The solid line is the 
linear fit for the set $(8,32)$ (left) and the set $(64,256)$ (right). The dash line is the behavior expected from equation (\ref{eq:linearfss}) and the accurate values of \ref{sec:appa}.}
\label{fig:fixedbet}
\end{figure}

\begin{figure}
\includegraphics[width=2.4in,angle=270]{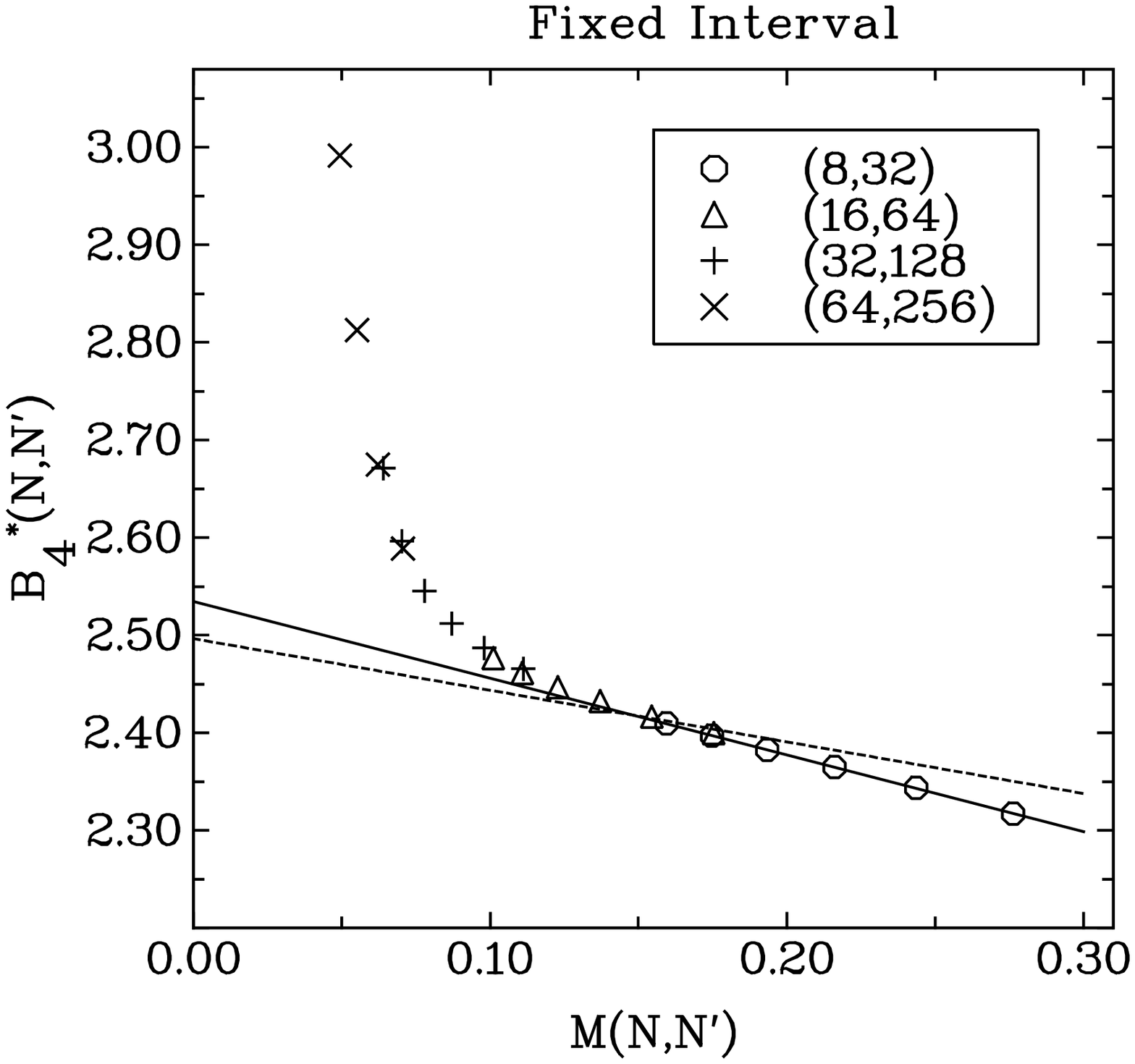}
\includegraphics[width=2.4in,angle=270]{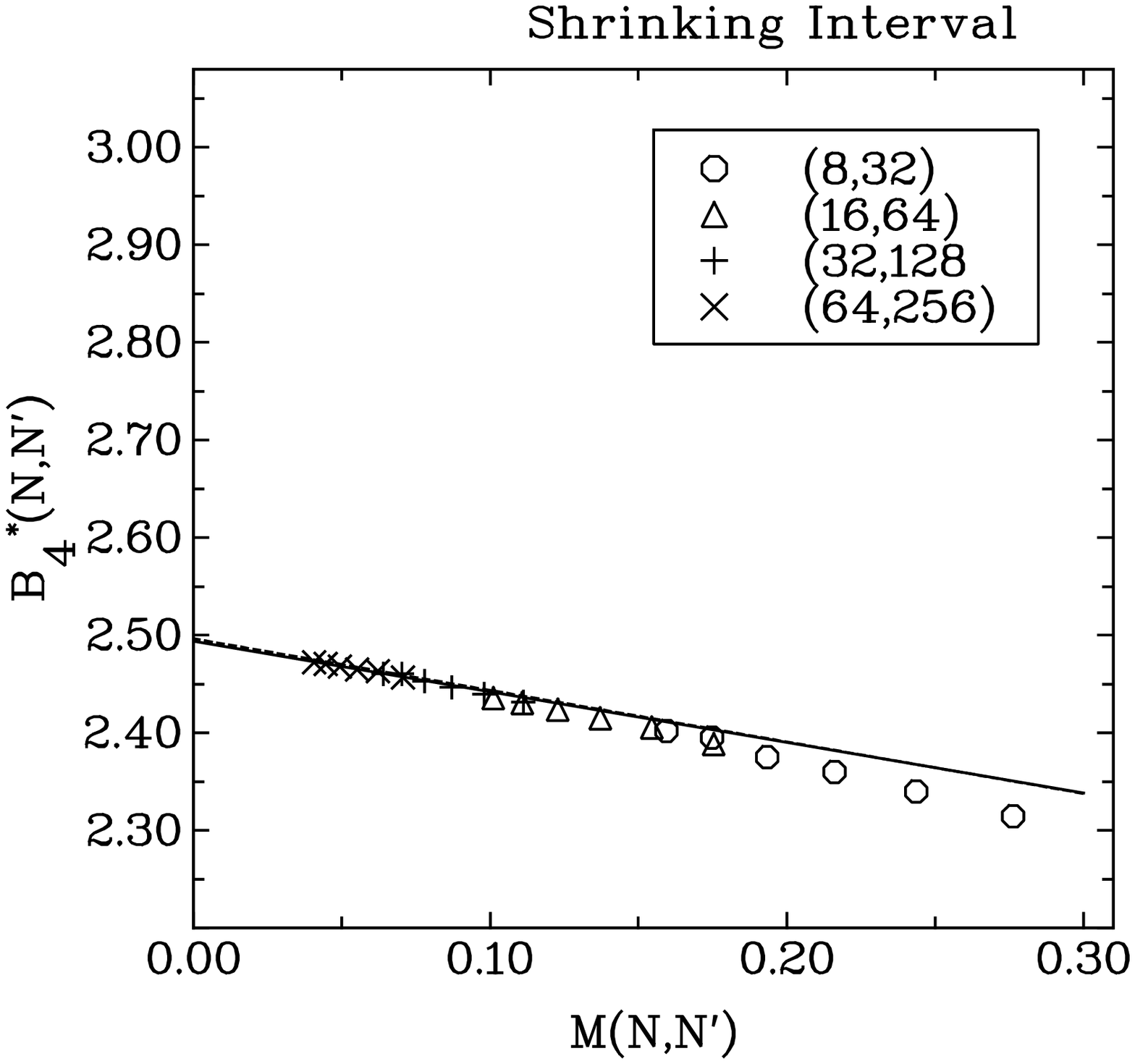}
\caption{Empirical values of $B_4^{\star}(N,N')$ versus $M(N,N')$ obtained with the fixed interval procedure (left) and with the shrinking interval procedure (right) for 4 sets of 6 pairs of values 
defined in the text. The solid line is the 
linear fit for the set $(8,32)$ (left) and the set $(64,256)$ (right). The dash line is the behavior expected from equation (\ref{eq:linearfss}) and the accurate values of section \ref{sec:appa}.}
\label{fig:fixedbin}
\end{figure}

Linear fits were performed for each $N$ in order to express $B_4$ 
as a linear function of $\beta$. For each pair $(N,N')$, it is possible to determine 
the intersection $(\beta^{\star}(N,N'),B_4^{\star}(N,N'))$ of the corresponding lines. 
These empirical values are plotted versus the calculable values $L(N,N')$ and $M(N,N')$, defined 
in (\ref{eq:LM}), in figures \ref{fig:fixedbet} and  \ref{fig:fixedbin}.
In order to make the graphs readable and the dependence on $N$ and $N'$ clear, we have displayed 4 sets of 
6 intersections denoted $(N,N')$ in the graphs. The meaning of $(N,N')$ is that we take the intersection between $N'$ and all the possible lower values starting at $N$. For instance, 
$(8,32)$ is a short notation for the 6 intersections: 8 and 32, $8\times 2^{1/3}$ and 32, ...., and, 
$8\times 2^{5/3}$ and 32. 

For the fixed interval procedure, one can see that values corresponding to the smallest volumes, namely 
the $(8,32)$ set, the behavior is approximately linear. The extrapolations at infinite volume 
for the fit with the six points of this set are 1.17848 for $\beta_c$ and 2.53435 for $B_4$, which 
is not too far from the accurate values but not very accurate either. However, if we increase the volume the linearity 
and the accuracy of the extrapolations degrade rapidly. 

On the other hand, with the shrinking interval method, the accuracy seems to improves with the size. In both graphs, the fit made with the $(64,256)$ set is hardly distinguishable from the accurate result. In order to give a more detailed idea of the effects on the extrapolated values, the numerical values 
of the parameters entering in equation (\ref{eq:linearfss}) are shown in table \ref{tab:shrin} for various sets of data. 
\begin{table}
\begin{center}
\begin{tabular}{||c|c|c|c|c||}
\hline
$n$& $\beta_c$ & $\beta_c c_0/f_1$&$B_4$&$c_0$\cr
\hline
15&1.1788071 & 3.74 & 2.5245& - 0.761\cr
16& 1.1789292 & 3.54 &2.51486& - 0.715\cr
17& 1.1789933 & 3.40 &2.50724& - 0.673\cr
18& 1.1790389 & 3.28 &2.49907& - 0.621\cr
19& 1.1790832 &2.99&2.48794 &- 0.534\cr
20& 1.1790042 & 3.51 &2.50940& - 0.697\cr
21& 1.1790198 & 3.35 & 2.50336& - 0.647\cr
22& 1.1790267 &3.20 &2.49945 &- 0.606\cr
23& 1.1790301 & 3.04 &2.49636& - 0.564\cr
24& 1.1790316 & 2.85 &2.49387& - 0.518\cr
\hline
Expected&1.1790302&2.91&2.49642&-0.529\cr
\hline
\end{tabular}
\end{center}
\caption{\label{tab:shrin} Values of $\beta_c$, $\beta_c c_0/f_1$, $B_4$ and $c_0$ obtained from linear fits with 
the $(2^{(n-6)/3},2^{n/3})$ set of 6 intersections described in the text.}
\end{table}

\begin{figure}
\includegraphics[width=3.4in,angle=270]{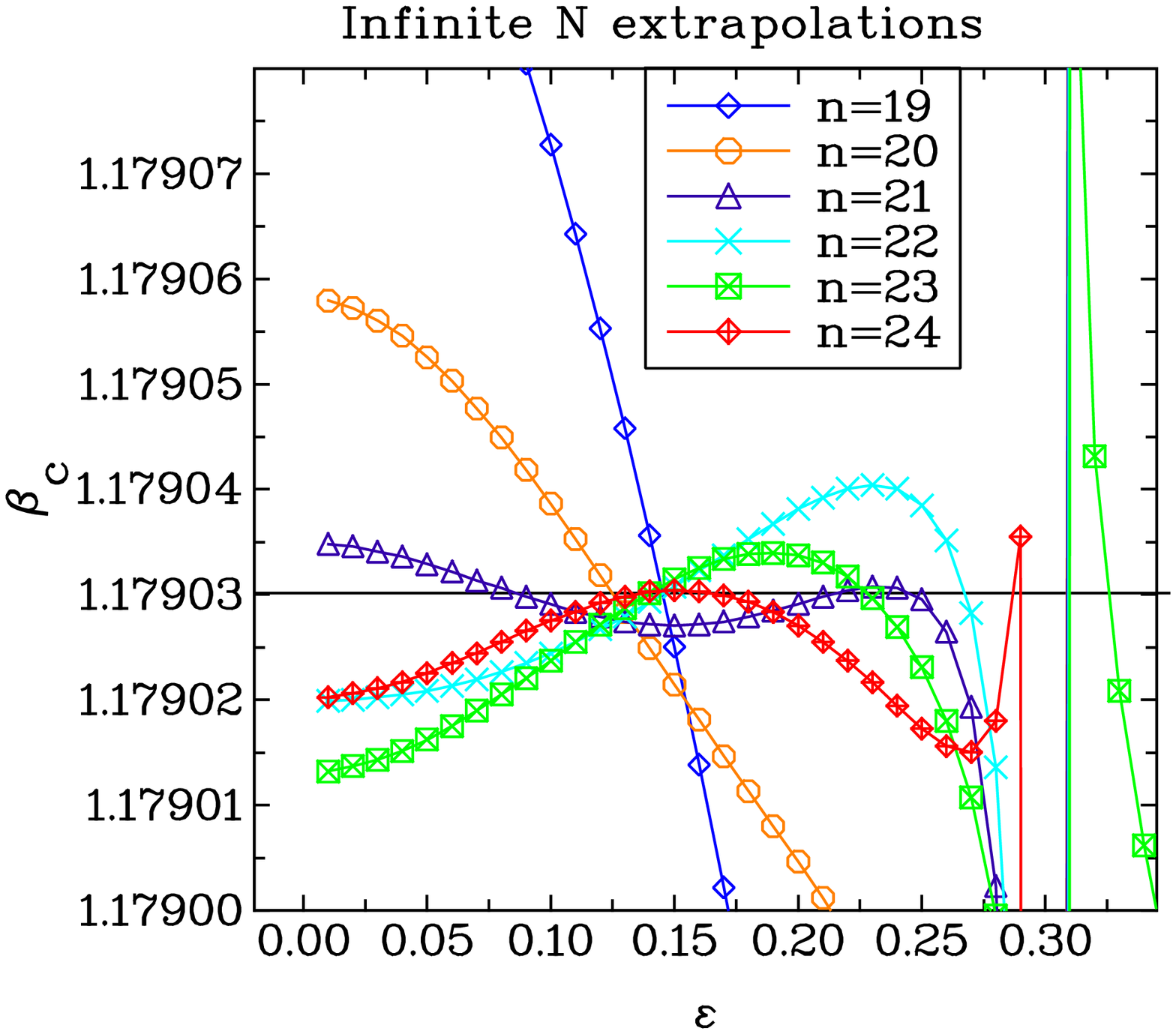}
\vskip5pt
\includegraphics[width=3.4in,angle=270]{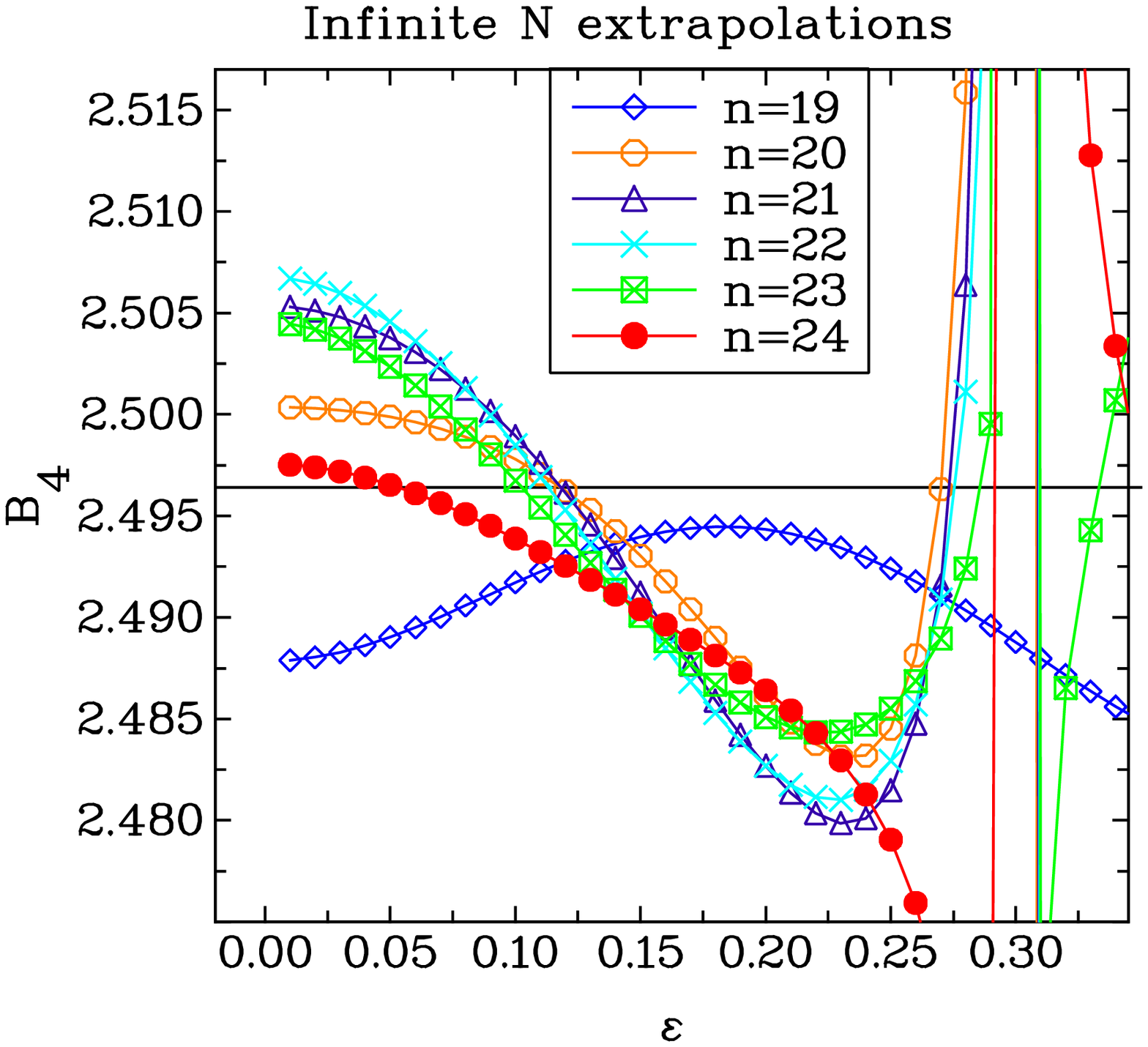}
\caption{Infinite volume extrapolations of $\beta_c$ and $B_4$ based on 15 point linear fits from the intersections among the $B_4$ curves at $N=2^{n/3}$ and the 5 values of $N$ immediately below, for $n$ between 19 and 24.}
\label{fig:extraps}
\end{figure}

We have also performed fits of the intersections for each value of $N$ using all the possible intersections 
involving the 5 values of the linear size immediately below. This amounts to 15 intersections, just as for the determination of $\bar{\beta_c}$. This procedure maximizes the number of data points for a given size interval and should be suitable when larger numerical errors are present.
We have repeated the calculation with smaller and larger values of $\epsilon$. The results for the $\beta_c$ and $B_4$ are shown in figure \ref{fig:extraps}. One sees that the results become erratic for $\epsilon>0.3$. There is a line crossing that 
appears in both graphs for $0.1<\epsilon< 0.15$. We believe that this crossing reflects the 
compromise discussed above, even though it occurs at a value of $\epsilon$ 
about two times larger than initially guessed. The estimates at the crossing are $\beta_c=1.179025(5)$ and 
$B_4=2.492(3)$. In general, we do not know precisely the crossing value and so larger error bars 
should be set by using the variation with $\epsilon$. From the figures, this should be 
at most $1\times 10^{-5}$ for $\beta_c$ and 0.005 for $B_4$ for $N=256$. The accurate values are clearly within the errors bar for the estimates obtained with the 6 and 15 points fits. It is also instructive to 
compare the 15 points fits in figure \ref{fig:extraps} at $\epsilon =0.05$ with the values of the 6 point fits given in table \ref{tab:shrin} for the same values of $n$. This gives an idea of the variability of the extrapolations as we change the data used to make the linear fits of equation (\ref{eq:linearfss}).

It is important to realize that near the value of $\epsilon$ where the various curves of figure \ref{fig:extraps} cross, the extrapolations are approximately the same for all the values of $n$.
This means that picking $\epsilon$ properly allows to get optimally accurate estimates with volumes 
not too large. One should also appreciate that there are 35 values of $\epsilon$ in each graph of 
figure \ref{fig:extraps} and that for each values of $\epsilon$, we need more than 100 calculations of 
$B_4$ to determine the intersections. Doing the same number of calculations using the MC method for spin models or lattice gauge theory 
would take a very long time. 
\section{Comments about the literature}
\label{sec:app}

In this section, we comment on existing numerical results found in the literature.
We first 
discuss existing estimates of $B_4$ for the Ising universality class. 
The methods used can be divided into those that rely on intersections \cite{ferrenberg91,gupta96}, and those that attempt to 
work as closely as possible to the nontrivial fixed point \cite{blote96,hasenbusch98,campostrini99}. 
Note that  \cite{blote96} takes into account the nonlinear effect of $f_2$ and also the effect of the second irrelevant direction. 
The results are summarized in 
table \ref{tab:ising}. The error bars in  \cite{ferrenberg91} are based on the visual estimate 
$U=0.47\pm 0.005$ from figure 7. The agreement between the two results based on improved action 
that reduce the effect of the irrelevant variables \cite{hasenbusch98,campostrini99} as well as the agreement with  \cite{blote96} indicates that 1.603 should 
be accurate with about one part in a 1000. The discrepancy with \cite{gupta96} seems to be of statistical origin: if we use the fit given in equation (14) of \cite{blote96} together with $N=256$ and $|K-K_c|<4\times 10^{-6}$ we obtain an absolute error less than $10^{-3}$ for $B_4$ if we neglect 
$f_2$. 
\begin{table}[t]
\begin{tabular}{||c|c|c|c|c||}
\hline
Model& $B_4$ & Ref. & Method & Intersections\cr
\hline
Ising 3 & 1.590 (15?)& \cite{ferrenberg91} & MC + reweighting& extrapolation\cr
\hline
Ising 3 & 1.591 (9)& \cite{gupta96} & MCRG + reweighting&(128,256)\cr
\hline
Ising 3 & 1.604(1)& \cite{blote96} & MC at best $\beta_c$ & -\cr
\hline
Ising 3 & 1.603(1)& \cite{hasenbusch98} & Improved action& -\cr
\hline
Ising 3 & 1.603(1)& \cite{campostrini99} & Improved action& -\cr
\hline
$SU(2)$ $N_\tau =4$ & 1.620(30) & \cite{fingberg92} & MC at various $\beta$& extrapolation \cr
$SU(2)$ $N_\tau =6$ & 1.570(80)& " & ''&"\cr
$SU(2)$ $N_\tau =8$ & 1.520(100)& " & ''&"\cr
\hline
$SU(2)$ $N_\tau =2$ & 1.622(3) & \cite{velytsky07} & MC at various $\beta$& (24,32)\cr
$SU(2)$ $N_\tau =4$ & 1.602(27)& " & ''&"\cr
$SU(2)$ $N_\tau =8$ & 1.584(39)& " & ''&"\cr
\hline
\end{tabular}
\caption{\label{tab:ising}Values of $B_4$ found in the literature for the two models discussed in the text. The last column provides additional information about a particular intersection $(N,N')$ used or about the fact that an extrapolation has been made.}
\end{table}

We now discuss the nonlinear effects for specific calculations made with the fixed interval procedure. 
For the 3 dimensional Ising model, $\beta_c$ (usually denoted $K_c$) is known 
with great accuracy and consequently, most of the graphs showing $B_4$ use a very 
narrow $\kappa$ interval where the differences between linear fits and numerical data are 
difficult to see. For instance, if we use the $N=16$ data in figure 6 of  \cite{barber85}, 
it is possible to draw a line that stays within the error bars of all the data points.  
Consequently, we can only set a lower bound on $f_1/f_2$ from this figure by assuming that $\Delta_0$, defined 
in \ref{sec:pencil}, is less than the error bars that we estimated to be less than 0.02. 
This bound is consistent with the numerical estimate of \cite{blote96}. 

A similar situation is encountered with 
figure 2 of  \cite{velytsky07} for $SU(2)$ LGT with $N_\tau=4$. We focus our analysis on the $N_\sigma=16$ 
data points. If we draw a line from the point with the lowest horizontal coordinate to the point 
with the largest one, the three points inside appear to be above the line. However, the line is within 
the three error bars. Using the data points kindly provided by the author, we obtained $\Delta_0 \simeq 
0.01$ which is slightly less than the errors bars. The lower bound quoted in the table is based on $\Delta_0 <0.015$. On the other hand, a clearly non-zero value for $\Delta_0$ can be seen in figure 11 
of  \cite{olsson96} and figure 9 of  \cite{engels89} and relatively stable values can be obtained 
for $f_1/f_2$. If we now denote $|\kappa N^{1/\nu}|_{max.}$ the maximum value of $|\kappa N^{1/\nu}|$ 
used in the linear fits made to determine the intersections (assuming that such a fits were performed, otherwise we rely on the width of the figure) in each reference, we can estimate the relative size of the nonlinear effects by calculating 
\begin{equation}
\epsilon_{max.}\equiv (f_2/f_1) |\kappa N^{1/\nu}|_{max.}\ .
\end{equation}
The numerical values are shown in table \ref{tab:emax}.
There are no large values of $\epsilon_{max.}$. 
This is because when $\epsilon_{max.}\sim 1$, the failure of linear fits is obvious. 
On the other hand, systematic errors due to unaccounted nonlinear effects may be 
reduced by using a smaller $\epsilon_{max.}$ with the procedure discussed in section \ref{sec:shrink}.
\begin{table}[t]
\begin{tabular}{||c|c|c|c|c||}
\hline
Model& Ising 3& Villain $XY$& $ SU(2)\ N_\tau=4$& $ SU(2)\ N_{\tau}=4$\cr
 \hline
 & \cite{barber85}&\cite{olsson96}&\cite{engels89}&\cite{velytsky07}\cr
 \hline
 Data used& fig. 6, $N=16$&Fit on fig. 11&fig. 9, $N_\sigma=8 $ &fig. 2, $N_\sigma=16$\cr
 \hline
 $f_1/f_2$&$>9(3)$&3&2&$>3(1.5)$\cr
 \hline
 $|\kappa N^{1/\nu}|_{max.}$ & 5.5& 0.4 &0.47& 0.4\cr
 \hline
 $\epsilon_{max.}$& $<0.6(2)$&0.13&0.24& $<0.13(6)$ \cr
 \hline
\end{tabular}

\caption{\label{tab:emax} Values of $f_1/f_2$, $|\kappa N^{1/\nu}|_{max.}$ and $\epsilon_{max.}$ for the four references discussed in the text.}
\end{table}
\section{Conclusions}
We have proposed a new method designed to reduce possible nonlinear effects in the estimates of $\beta_c$ and $B_4$. For the Ising hierarchical model with a volume corresponding to a linear size of 
256 in 3 dimensions, the method gives results which agree with independent accurate estimates with 
a relative accuracy better than one part in 100,000 for $\beta_c$ and 5 parts in 1000 for $B_4$. 
The intrinsic nonlinear effects (measured in terms of $f_2/f_1$) in this model are roughly of the same size as those found the other models 
that we discussed in section \ref{sec:app}. Figures \ref{fig:fixedbet} and \ref{fig:fixedbin} shows the potentially disastrous effect of ignoring nonlinear effects. In graphs with large numerical errors, this effect may be overlooked. The small values of $\epsilon_{max}$ found in the literature 
indicate that nothing drastic should appear in the cases considered.
However, it would be interesting 
to repeat and extend these calculations for larger volumes using the method proposed here. 
The optimal value of $\epsilon$ can in principle be obtained by looking at crossings of extrapolated values. This suggests that the procedure should be followed for a few different values of $\epsilon$. 
We are planning to consider the effects of statistical errors on the new method by applying it to 
the 3 dimensional Ising model and finite temperature gluodynamics.  
\ack
We thank A. Velytsky for valuable discussions and comments and for providing his numerical data. 
This 
research was supported in part  by the Department of Energy
under Contract No. FG02-91ER40664.
\appendix
\section{Numerical Calculations}
\label{sec:appa}
For completeness, we explain how we calculated $B_4$ for the hierarchical model and we give the details of the numerical calculations of the parameters.  
The interest of the hierarchical model is that we can blockspin exactly. 
The basic formula for the Fourier transform of the local measure is 
\begin{equation}
R_{n+1}(k) = C_{n+1}
{\rm e}
^{-(1/2)\beta (c/4)^{n+1}(\partial^2/\partial k^2)}
R^2_n(k)\ ,
\label{eq:rec}
\end{equation}
with
\begin{equation}
\label{eq:cdim}
	c=2^{1-2/D}\ .
\end{equation}
Note that we do not rescale the field as in a renormalization group transformation. 
$C_{n+1}$ is a constant that we adjust so that $R_{n+1}(0)=1$. When it is the case, we have 
\begin{equation}
R_n(k) =1+ \sum_{q=2}^{\infty}{(-k)^{2q}/(2q)!}<(\phi_n)^{2q}>_n \ ,
\label{eq:gen}\end{equation}
with $\phi_n$ the sum of all the spins. 
If we write
\begin{equation}
R_n(k) = 1 + a_{n,1}k^2 + a_{n,2}k^4 + ... + a_{n,l_{max}}k^{2l_{max}}\ .
\end{equation}
Then 
\begin{equation}
B_4=6a_{n,2}/a_{n,1}^2\ .
\end{equation}
The critical values are obtained by plugging the values of $a_1^{\star}$ and $a_2^{\star}$ from the 
nontrivial fixed point. See  \cite{hmreview} for details and references. 
The value of $c_0+c_1(\kappa N^{1/\nu})$ is obtained by increasing $N$ but keeping $\kappa N^{1/\nu}$ 
constant. The asymptotic values for different $\kappa N^{1/\nu}$  can be fitted very well with a line which gives 
$c_0$ and $c_1$. We then subtracted the effect of $c_0+c_1(\kappa N^{1/\nu})$ and obtained $f_1$ and $f_2$ using discrete derivatives near $\kappa =0$.

We used calculations at very large volume ($N\sim 10^6$). In the text, we show that 
these accurate results can be reproduced with a new procedure using data at much smaller volume.
From Refs. \cite{gam3rapid,gam3},
\begin{eqnarray}
\nu &=&0.649570365\ ,\nonumber \\
\omega &=&0.655736\ ,\nonumber  \\
B_4(\beta_c,\infty) &=&2.49641845\ ,\\
\beta_c &=&1.17903017044626973251\ .\nonumber
\end{eqnarray}
Using these values, and numerical results a fixed values of $\kappa N^{1/\nu}$ and large $N$, we first 
determine the $N^{-\omega}$ terms and find 
\begin{eqnarray}
	c_0&=&-0.529\ , \nonumber\\
	c_1&=&- 0.236\ .
\end{eqnarray}
Subtracting these effect and taking discrete approximation of the derivative with respect to $\kappa$ near 0,  we find 
\begin{eqnarray}
	f_1&=& -0.214\ , \nonumber \\
	f_2&=&-0.051\ .
\end{eqnarray}
One should appreciate that we have been able to determine numerically the 8 parameters of 
equation (\ref{eq:param}) with good numerical stability.
\section{Estimate of the nonlinear effects with pencil and ruler}
\label{sec:pencil}
\begin{figure}
\includegraphics[width=2.4in,angle=270]{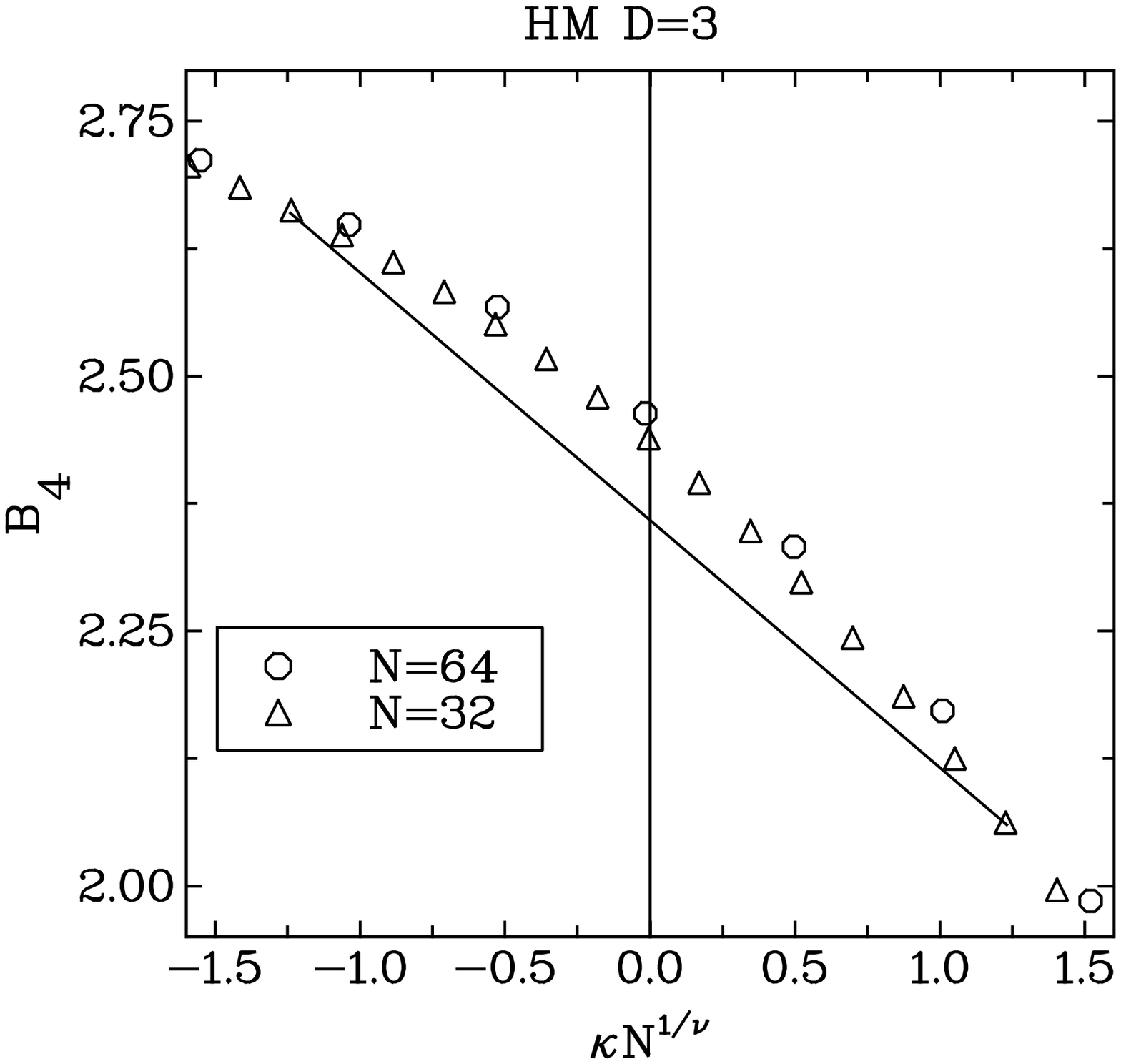}
\includegraphics[width=2.4in,angle=270]{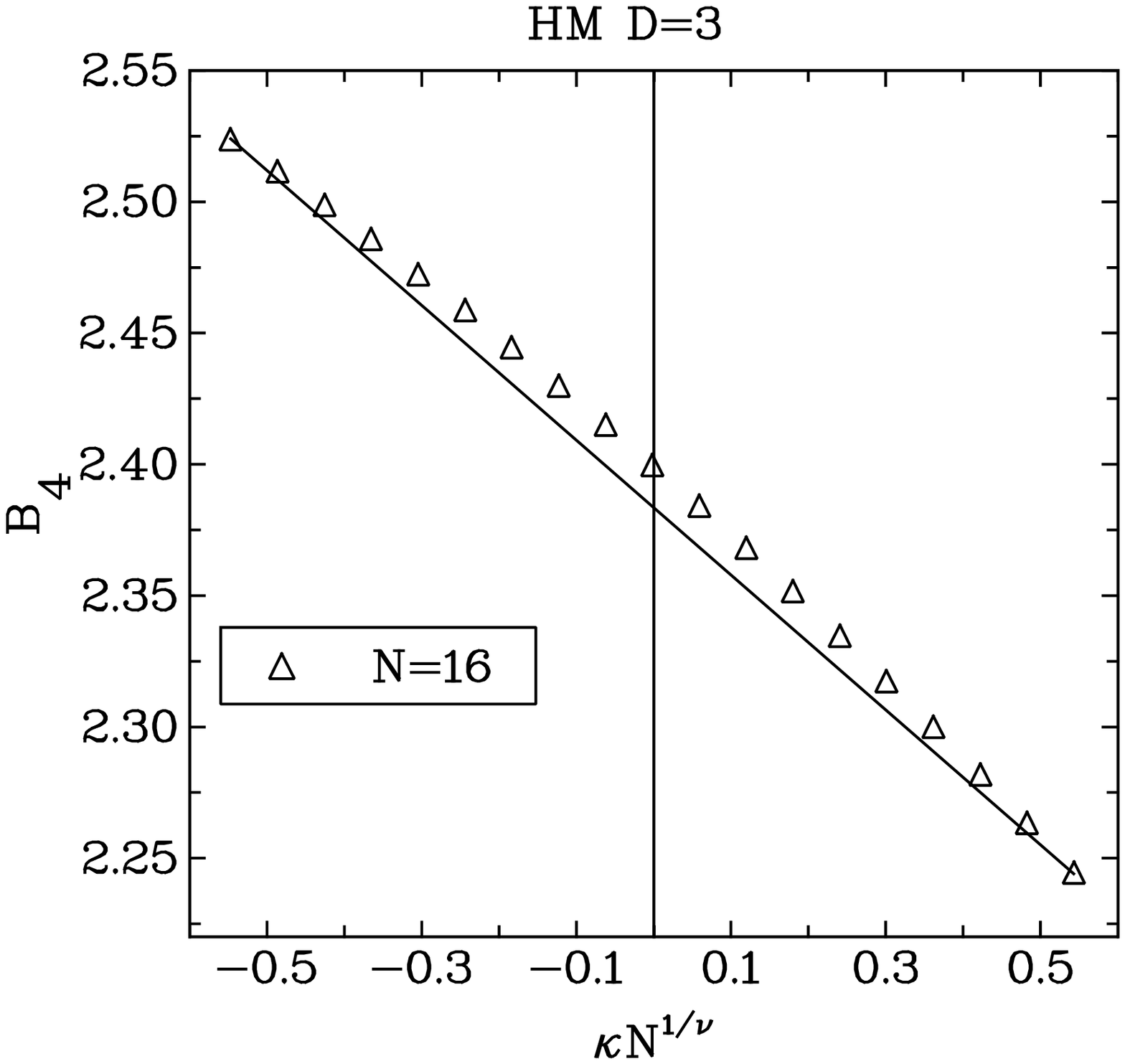}
\caption{$B_4$ versus $\kappa N^{1/\nu}$, for $N=$ 16 (below), 32 and 64 (above) for the Ising hierarchical  model.}
\label{fig:penrule}
\end{figure}

In this appendix, we explain how to estimate the order of magnitude of  $f_1/f_2$ from a graph illustrating approximate data collapse such as figure \ref{fig:b4}, using a pencil and a ruler. A subset of the data of figure \ref{fig:b4} is shown 
in figure \ref{fig:penrule} together with lines that can be drawn on the original graph. 
The line joins two points with approximately opposite $x$ coordinates. We denote these two points 
$(x,y_1)$ and $(-x,y_2)$. We call $\Delta_0$ the difference between the line and the data (assumed to be a 
quadratic function) at 0 . It is then easy to show that 
\begin{eqnarray}
	f_1&\simeq&(y_1-y_2)/2x \nonumber \\
	f_2&\simeq&-\Delta_0/x^2
\end{eqnarray}
In figure \ref{fig:penrule}, we have approximately $x\simeq 1.24$, $y_2\simeq2.66$, $y_1\simeq 2.06$ and $\Delta_0\simeq 0.07$, which gives the estimates $f_1\simeq -0.24$ and $f_2\simeq -0.045$ which are in 
reasonably 
good agreement with the previous estimates. 
Note that we are mostly interested in the ratio $|f_1/f_2|$ and that this ratio is invariant under a 
multiplicative rescaling. This is useful if we are considering graphs with other quantities displayed 
(for instance $1-B_4/3$). For the model considered here, $|f_1/f_2|\simeq 5$.

The above estimates are based on the data with $N=32$. In figure \ref{fig:penrule}, we also see part of 
the data for $N=64$ and we see that it is not much  above the $N=32$ data. From equation (\ref{eq:param}), it is clear that if we neglect $c_1$, the entire curve is translated uniformly in the vertical direction. 
But the estimates of $f_1$ and $f_2$ are based on differences and do not depend on this translation provided that we only use data for one value of $N$. This is confirmed by using the $N=16$ data as shown in figure \ref{fig:penrule} where we obtain an estimate of $f_1/f_2$ very close to the one quoted above for $N=32$. This shows that up to a subtraction and rescaling of the horizontal coordinate, we can rely 
on a graph of $B_4$ versus $\beta$ at relatively small volume. It is clear that the subtraction of $\beta_c$ requires to have an estimate of this quantity. However, this estimate does not need to be 
very precise to get the order of magnitude of $f_1/f_2$.  
\vskip15pt
\noindent
{\bf References}
\vskip10pt

\begin{thebibliography}{10}
\expandafter\ifx\csname url\endcsname\relax
  \def\url#1{{\tt #1}}\fi
\expandafter\ifx\csname urlprefix\endcsname\relax\def\urlprefix{URL }\fi
\providecommand{\eprint}[2][]{\url{#2}}

\bibitem{binder81}
Binder K 1981 {\em Z. Phys.\/} {\bf B43} 119--140

\bibitem{barber85}
Barber M~N, Pearson R~B, Toussaint D and Richardson J~L 1985 {\em Phys. Rev.
  B\/} {\bf 32} 1720--1730

\bibitem{ferrenberg91}
Ferrenberg A~M and Landau D~P 1991 {\em Phys. Rev. B\/} {\bf 44} 5081--5091

\bibitem{olsson96}
Olsson P 1997 {\em Phys. Rev. B\/} {\bf 55} 3585--3602

\bibitem{blote96}
Blote H, Luitgen E and Heringa J 1996 {\em J. Phys. A\/} {\bf 28} 6289--6313

\bibitem{gupta96}
Gupta R and Tamayo P 1996 {\em Int. Jour. of Mod. Phys. C\/} {\bf 7} 305--319

\bibitem{hasenbusch98}
Hasenbusch M, Pinn K and Vinti S 1998 
{\em Physical Review B \/} {\bf 59} 11471 

\bibitem{campostrini99}
Campostrini M, Pelissetto A, Rossi P and Vicari E 1999 {\em Phys. Rev.\/} {\bf
  E60} 3526--3563 

\bibitem{binder01}
Binder K and Luijten E 2001 {\em Phys. Rept.\/} {\bf 344} 179--253

\bibitem{engels89}
Engels J, Fingberg J and Weber M 1990 {\em Nucl. Phys.\/} {\bf B332} 737

\bibitem{fingberg92}
Fingberg J, Heller U~M and Karsch F 1993 {\em Nucl. Phys.\/} {\bf B392}
  493--517 

\bibitem{sinclair06}
Sinclair D~K and Kogut J~B 2006 {\em PoS\/} {\bf LAT2006} 147
  (\textit{Preprint} \eprint{hep-lat/0609041})

\bibitem{sinclair07}
Sinclair D~K and Kogut J~B 2007  (\textit{Preprint} \eprint{arXiv:0709.2367
  [hep-lat]})

\bibitem{deforcrand07a}
de~Forcrand P, Stephanov M~A and Wenger U 2007  (\textit{Preprint}
  \eprint{arXiv:0711.0023 [hep-lat]})

\bibitem{deforcrand07b}
de~Forcrand P, Kim S and Philipsen O 2007  (\textit{Preprint}
  \eprint{arXiv:0711.0262 [hep-lat]})

\bibitem{velytsky07}
Velytsky A 2007  (\textit{Preprint} \eprint{arXiv:0711.0748 [hep-lat]})

\bibitem{moprogress}
Meurice Y and Oktay B in progress

\bibitem{dyson69}
Dyson F 1969 {\em Comm.\ Math.\ Phys.\/} {\bf 12} 91

\bibitem{guide}
Godina J, Meurice Y, Oktay M and Niermann S 1998 {\em Phys. Rev. D\/} {\bf 57}
  6326

\bibitem{gam3rapid}
Godina J, Meurice Y and Oktay M 1998 {\em Phys. Rev. D\/} {\bf 57} R6581

\bibitem{gam3}
Godina J, Meurice Y and Oktay M 1999 {\em Phys. Rev. D\/} {\bf 59} 096002

\bibitem{hmreview}
Meurice Y 2007 {\em J. Phys.\/} {\bf A40} R39 

\bibitem{campbell06}
Campbell I~A, Hukushima K and Takayama H 2006 {\em Physical Review Letters\/}
  {\bf 97} 117202

\bibitem{campbell07}
Campbell I~A, Hukushima K and Takayama H 2007 {\em Physical Review B \/} {\bf 76} 134421 
\end{thebibliography}
\providecommand{\newblock}{}

\end{document}